\documentclass[aps,prb,twocolumn,showpacs,preprintnumbers,amsmath,amssymb,superscriptaddress]{revtex4}

\usepackage{amssymb}
\usepackage{amsmath}
\usepackage{fancybox}

\usepackage[dvips]{color}
\usepackage{epic}

\usepackage{graphicx}
\usepackage{dcolumn}
\usepackage{bm}
\newcommand{\nc}{\newcommand}
\nc{\be}{\begin{equation}}
\nc{\ee}{\end{equation}}
\nc{\bea}{\begin{eqnarray}}
\nc{\eea}{\end{eqnarray}}
\nc{\bean}{\begin{eqnarray*}}
\nc{\eean}{\end{eqnarray*}}
\nc{\mb}{\mbox}
\nc{\rnc}{\renewcommand}
\nc{\vk}{\mb{\bf k}}
\nc{\vp}{\mb{\bf p}}
\nc{\vn}{\mb{\bf n}}
\nc{\vq}{\mb{\bf q}}
\nc{\rr}{\mb{\bf r}}
\nc{\vz}{\hat {\mb{\bf z}}}
\nc{\vj}{\mb{\boldmath$j$}}
\nc{\vg}{\mb{\boldmath$g$}}
\nc{\x}{\mb{\boldmath$x$}}
\nc{\A}{\mb{\boldmath$A$}}
\nc{\va}{\mb{\boldmath$a$}}
\nc{\vs}{\mb{\boldmath$\sigma$}}
\nc{\vpi}{\mb{\boldmath$\pi$}}
\nc{\nab}{\nabla}
\nc{\X}{\sf x}

\begin{document}

\title{
Fractional Quantum Hall Effects in Graphene and Its Bilayer
}
\author{Naokazu Shibata and Kentaro Nomura}
\affiliation{Department of Physics, Tohoku University, Aoba, Aoba-ku,
Sendai, 980-8578, Japan} 

\date{June 5, 2009}

\begin{abstract}
Single-layer and Bilayer of graphene are new classes of two-dimensional
electron systems with unconventional band structures and valley degrees
of freedom. 
The ground states and excitations in the integer and fractional quantum
Hall regimes are investigated on torus and spherical geometries with
the use of the density matrix renormalization group (DMRG) method. 
At nonzero Landau level indices, the ground states at effective filling
factors $1$, 1/3, 2/3 and 2/5 are valley polarized both in single-layer and
bilayer graphenes.  
We examine the elementary charge excitations which could couple with
the valley degrees of freedom (so called valley skyrmions).  
The excitation gaps are calculated and extrapolated to the thermodynamic
limit.
The largest excitation gap at effective  
filling $1/3$ is obtained in bilayer graphene, which
is a good candidate for experimental
observation of fractional quantum Hall effect.
\end{abstract}

\maketitle

\section{Introduction}

A recent experimental realization of single-layer graphene (SLG) sheets
\cite{Novoselov_2004} has made it possible to confirm a number
of theoretical predictions of intriguing electric properties of  
 massless Dirac fermion systems,\cite{Neto_2007}
including unconventional quantum Hall effects
(QHE)\cite{Novoselov_2005,Zhang_2005} 
with the half-integer Hall conductivity\cite{Dirac_qhe} 
\bea
\sigma_{xy} =  (n +\frac{1}{2})\frac{4e^2}{h},
\eea
at $\nu=\pm2,\pm6,\pm10, \cdots$, where a factor $4$ is the Landau level (LL) 
degeneracy, resulting from spin and valley (referred to K and K')
symmetry in graphene. 
$\nu=2\pi\ell_B^2\rho$ is the filling factor, $\ell_B=\sqrt{\hbar/eB}$
is the magnetic length, $\rho$ is the carrier density measured from
the charge neutral Dirac point. 

As many phenomenological insights were found 
in the quantum Hall regime of SLG
\cite{Novoselov_2005,Zhang_2005,Dirac_qhe,Zhang_2006,Nomura_2006,
Alicea_2006,Yang_2006,Goerbig_2006,Chakraborty_2007,Abanin_2007,
Sheng_2007,Apalkov_2006,Shibata_2008},
bilayer graphene (BLG) potentially exhibits rich
physics.\cite{Novoselov_2006,McCann_2006} Experiments showed that the
Hall conductivity of unbiased graphene bilayer in strong magnetic 
fields is given by 
\bea
\sigma_{xy} =  n\ \frac{4e^2}{h}
\eea
with $|n|\ge 1$.\cite{Novoselov_2006} 
The absence of $\sigma_{xy}=0$ plateau and 
the double height jump of the Hall conductivity between $\nu=-4$ and $4$
indicate eight-fold degeneracy at the neutrality point.  

These quantization rules of the Hall conductivity for SLG and BLG
originate with the characteristic energy spectrum in a magnetic field. 
The low-energy band structure in SLG consists of Dirac cones located at
the inequivalent Brillouin zone corners K and K'. Using the magnetic
ladder operator $a\equiv (\pi_x-i\pi_y)\ell_B/\sqrt{2}\hbar$, where
${\bf \pi}={\bf p}-e{\bf A}$ is the kinetic momentum, 
the single-particle Hamiltonian for the K-valley of the SLG is written as
\bea
{\cal H}_{\rm K}^{\; \rm (SLG)}  = \ \ 
\frac{\sqrt{2}\hbar v_F^{}}{\ell_B}
\left(
\begin{array}{rr}
 0\ \   &  a   \\
 a^{\dag}  & 0  
\end{array}
\right)\; ,
\label{hamiltonian-slg}
\eea
and ${\cal H}_{K'}^{\rm (SLG)}$ is the transpose of ${\cal H}_{K}^{\rm (SLG)}$.
Where $v_F$ is the Fermi velocity of SLG, 2$\times$2 matrices act in the
sublattice degrees of freedom in graphene. 
The eigenenergy of eq.(\ref{hamiltonian-slg}) is given by 
\bea
\epsilon_{n}=\pm\hbar v_F^{}\sqrt{2n}/\ell_B.
\eea
The eigenvector is 
$|0\rangle_{K}^{\rm (SLG)}=(0,|0\rangle)^t$ 
for $n=0$, and 
$|n,\pm\rangle_{K}^{\rm (SLG)}=(|n-1\rangle,\pm|n\rangle)^t/\sqrt{2}$
for $n\ge 1$.
Here $|n\rangle$ is the eigenvector of the number operator $N\equiv
a^{\dag}a^{}$ with an eigenvalue $n$. 

In contrast to SLG, BLG has an ordinary parabolic spectrum in
the vicinity of the neutrality point. In a magnetic field, 
the effective Hamiltonian for BLG is written in the form:
\cite{McCann_2006}
\bea
{\cal H}_{\rm K}^{\; \rm (BLG)}  = \ \ 
\hbar\omega_c
\left(
\begin{array}{rr}
  0\ \ \   &   a^2_{\;}   \\
 (a^{\dag})^2  & 0\    \
\end{array}
\right)\; ,
\label{hamiltonian-blg}
\eea
and its eigenenergy is 
\bea
E_{n}=\pm \hbar\omega_c\sqrt{n(n-1)}.
\eea
Here $\omega_c=eB/m$ with $m$ being the effective mass of BLG.
The eigenvector of eq.(\ref{hamiltonian-blg}) is
$|0,0\rangle_{K}^{\rm (BLG)}=(0,|0\rangle)^t$, 
$|0,1\rangle_{K}^{\rm (BLG)}=(0,|1\rangle)^t$ 
for the zero energy level, 
and 
$|n,\pm\rangle_{K}^{\rm (BLG)}=(|n-2\rangle,\pm|n\rangle)^t/\sqrt{2}$
for $n\ge 2$.

Recent experimental studies on SLG in a sufficiently strong magnetic field 
revealed new quantum Hall states at $\nu=0,\pm1,\pm4$,\cite{Zhang_2006} 
where the electron-electron interaction may play a crucial role. 
%
Here relevant energy scales in graphene in a magnetic field are   

(i) LL separation around the neutrality point, 
$\sqrt{2}\hbar v_F^{}/\ell_B 
\simeq 400\sqrt{B[T]}[K]$ for SLG, while $\sqrt{2}\hbar\omega_c\simeq
30\times (B{\rm [T]}) {\rm [K]}$ for BLG. 

(ii) Zeeman coupling, $\Delta_z\equiv g\mu_B|{\bf B}|\simeq
1.5\times(B[T])[K]$, and  

(iii) The Coulomb energy, $e^2/\epsilon \ell_B\simeq 100\sqrt{B[T]}[K]$.

The activation energy measurements at these additional QHE states in
SLG\cite{Zhang_2006} have shown that at 
$\nu=\pm4$ the gap has linear $B$ dependence and reasonably corresponds 
to $\Delta_z$, indicating Zeeman spin splitting. 
At $\nu=\pm 1$, on the other hand, the gap is approximately scaled by
$\sqrt{B}$,  
that indicates the gap originates from the Coulomb interaction. 
The latter behavior is consistent with the quantum Hall
ferromagnetism (QHF)\cite{QHF_review} 
in which valley degrees of freedom, referred as pseudospins,
spontaneously split via the exchange energy at {\it all integer} 
fillings.\cite{Nomura_2006,Alicea_2006,Goerbig_2006,Yang_2006,
Chakraborty_2007,Sheng_2007,Abanin_2007}   
Based on the above scenario, it has been anticipated that interactions
will drive quantum Hall effects also in BLG, 
at the octet's seven intermediate
integer filling factors when magnetic field is 
strong enough or disorder is weak enough.\cite{Barlas_2008} 

In this work, we focus on the many-body states in nonzero LLs 
of SLG and BLG.
We introduce the effective filling factors 
of the topmost partially filled $n$th LL,
\bea
 {\nu}_{n}^{\; \rm (SLG)}=\nu-4(n-1/2)
\eea
for SLG and
\bea
 {\nu}_{n}^{\; \rm (BLG)}=\nu-4(n-1)
\eea
for BLG. Here ${\nu}_{n} \le 4$, and $n\ge 2$ for BLG.

The ground state at $\nu_n=1$ is fully spin and valley polarized, and
the wave function can be represented by  
\bea
 |\Psi^{{ {\nu}_n}=1}_{\tau}\rangle
=\prod_{m}c^{\dag}_{m,\tau}|0\rangle, 
\eea 
where we assign the valley $K$ and $K'$ in graphene
as $z$-component of pseudospin 
$\tau=K$ or $K'$. We omit real spins, which are assumed to be fully
polarized by the strong Zeeman splitting. 
Excitations from the symmetry broken states are described by
(pseudo)spin wave and (pseudo)spin textures called 
skyrmions\cite{Sondhi_1993,Fertig_1994,Moon_1995,Rezayi_1991,Yoshioka_1998}
or other types, depending on the LL index $n$.\cite{Yang_2006}

In the previous work, we have studied the fractional quantum Hall 
ferromagnetic 
states in SLG at $\nu_n=1$ and $1/3$.\cite{Shibata_2008}
As a consequence of the relativistic nature of electrons in SLG, the
effective electron-electron interactions in $n\ne 0$ LLs differ from that
in conventional two-dimensional systems, while the $n=0$ LL is
equivalent.\cite{Nomura_2006,Goerbig_2006,Apalkov_2006} 
We have neglected the spin degrees of freedom of electrons 
by assuming strong Zeeman coupling, although
valley degrees of freedom have been taken into account 
since there is no external symmetry breaking field.
We have shown that the ground states at $\nu_n=1$ and $1/3$
in $n=0$ and 1 LLs of SLG 
are fully valley polarized, while elementary charge excitations consist of 
pseudospin-singlet, namely valley-skyrmions.\cite{Sondhi_1993,Fertig_1994,Moon_1995,Rezayi_1991,Yoshioka_1998}  

In this paper we extend the quantum Hall ferromagnetism to BLG and
numerically show that the ground states at $\nu_n=1$ 
and $1/3$ in the $n=2$ LL of BLG are fully valley polarized.
In contrast to the fact that elementary charge excited states 
are pseudospin-singlet at $\nu_2=1$, those
at $\nu_2=1/3$ of BLG are pseudospin
polarized. Namely, the Laughlin type quasiparticle (quasihole)
excitations dominate over skyrmion type. 
As a consequence, the charge gap at $\nu_2=1/3$ in BLG is 
almost twice larger than that at $\nu_0=1/3$ in SLG. 
Therefore clean BLG
samples are better candidates to observe fractional QHE than SLG. 
We also study $\nu_n=2/3$ and 2/5 states in SLG and BLG.
We claim that the ground at $\nu_n=2/3$ and 2/5 
states are pseudospin singlet in the $n=0$ LL of SLG, while 
fully pseudospin polarized in the $n=1$ LL of SLG
and the $n=2$ LL of BLG.
These results are summarized in Table I.

\section{model and method}

We start with the projected Coulomb interaction Hamiltonian onto a
certain LL, and study charge and valley excitations,
where we treat the valley degrees of freedom  $K$ and $K'$ in the
language of the pseudospin, while real spin degrees of freedom
are supposed to be frozen by the Zeeman splitting. 
The LL mixing is also neglected in the following.
We calculate the exact wave function of the ground state and low energy
excited states in SLG and BLG, basing on the density matrix
renormalization group (DMRG) method,\cite{white,shibata1,shibata2}
and examine the existence of the skyrmion excitations in 
valley degrees of freedom not only at integer fillings $\nu_n=1$ 
but also at fractional fillings $\nu_n=1/3$, 2/5 and 2/3 for the 
LL indices $n=0$, 1 and 2.  
Charge gaps at these fractions are extrapolated to the thermodynamic limit. 
For our purpose the DMRG method is quite useful, since
it needs to treat a large number of basis of the many-body Hilbert space
when the pseudospin degrees of freedom are introduced in the 
fractional QHE systems.  
Taking account of valley degrees of freedom, 
we apply the DMRG method\cite{white} on 
torus\cite{shibata1,shibata2} and spherical 
geometries.\cite{Haldane_1983,QHE_text,Fano_1986,Morf_2002,
Feiguin_2007,Shibata_2008}

The projected Hamiltonian onto the $n$th LL is written
as\cite{Haldane_1983} 
\bea
 H=\frac{1}{L^2}\sum_{i<j}\sum_{\bf q}V(q)e^{-q^2/2}[F_n(q)]^2
e^{i{\bf q}\cdot{({\bf R}_i-{\bf R}_j)}},
\label{hamiltonian1}
\eea
where ${\bf R}_i$ is the guiding center coordinate of the 
$i$th particle.
The {\it relativistic} form factor in the $n$th LL
of SLG is written as \cite{Nomura_2006}
\bea 
 F_0^{\rm (SLG)}(q)=L_0\left({q^2}/{2}\right)
\eea 
and 
\bea
 F_{n\ge 1}^{\rm (SLG)}(q)=(1/2)\left[L_{|n|}\left({q^2}/{2}\right)
+L_{|n|-1}\left({q^2}/{2}\right)
\right].
\eea
For the $n(\ge2)$th LL in BLG,
\bea
 F_{n\ge 2}^{\rm (BLG)}(q)=(1/2)\left[L_{|n|}\left({q^2}/{2}\right)
+L_{|n|-2}\left({q^2}/{2}\right)
\right].
\eea
Here $L_n(x)$ are the Laguerre polynomials. 
In the spherical geometry, it is convenient to write the Hamiltonian as
\bea
 H^{(n)}=\sum_{i<j}\sum_{m}V^{(n)}_mP_{ij}[m],
\label{hamiltonian2}
\eea
where $P_{i,j}[m]$ projects onto states in which particles $i$ and $j$
have relative angular momentum $\hbar m$, and $V^{(n)}_m$ is their
interaction energy in the $n$th LL.\cite{Haldane_1983}
Using above form factors, the
pseudopotentials\cite{QHE_text,Haldane_1983} are given by 
\bea
V^{(n)}_m=\int_0^{\infty}\frac{dq}{2\pi}qV(q)e^{-q^2}[F_n(q)]^2L_m(q^2).
\eea
The corresponding integrals
for electrons on the surface of a sphere which are used in the
present work are described in
refs.~\onlinecite{Haldane_1983,Fano_1986,Morf_2002}.  
Note that our Hamiltonian eqs.~(\ref{hamiltonian1}) and
(\ref{hamiltonian2}) have SU(2) symmetry in 
the valley degrees of freedom.  
A symmetry breaking correction to eqs.~(\ref{hamiltonian1}) and
(\ref{hamiltonian2}) which stems 
from the lattice structure of
SLG\cite{Goerbig_2006,Alicea_2006} and BLG is order of $a/\ell_B$ ($a$
being a lattice spacing) in units of $e^2/\epsilon \ell_B$ and neglected
in the following. 

We calculate the ground state wave function using the DMRG
method,\cite{white,shibata1} which 
is a real space renormalization group method combined with the exact
diagonalization method. The DMRG method provides low-energy
eigenvalues and corresponding eigenvectors of the Hamiltonian within a
restricted number of basis states. 
The accuracy of the results is systematically controlled by the
truncation error, which is smaller than $10^{-4}$ in the present
calculation. We investigate systems of various sizes with up to 40
electrons in the unit cell keeping 1400 basis in each
block.\cite{shibata1,shibata2}

\begin{figure}[t]
\begin{center}
\includegraphics[width=0.47\textwidth]{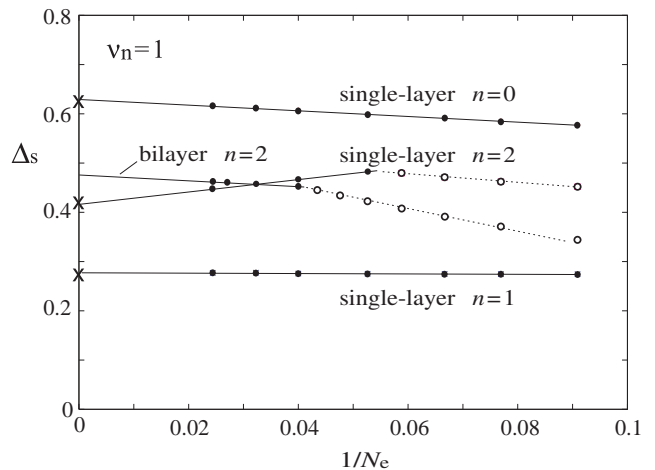}
\caption{The lowest charge excitation gaps
in SLG and BLG at $\nu_n=1$ in the $n=0$, 1 and 2
 LLs in the spherical geometry. 
Closed circles represent the pseudospin unpolarized $P=0$
(skyrmion) excitation gaps and open circles
represent nearly pseudospin-polarized $P=1-2/N_e$
excitation gaps.
The crosses on the vertical axis represent results 
obtained by Hartree-Fock calculations.\cite{Yang_2006}
}
\label{figure1}
\end{center}
\end{figure}

The sphere geometry is useful to extrapolate energy gaps
to the thermodynamic limit. 
In the sphere geometry, 
the pseudospin (valley) polarized 
ground state at $\nu_n=1/q$ ($q$ being an odd integer), 
the Laughlin state,\cite{QHE_text} realizes 
when the total flux $N_{\phi}$ is given
by\cite{Haldane_1983} 
\bea
 N_{\phi}(\nu_n,N_e)=\nu_n^{-1}(N_e-1),
\eea
where $N_e$ is the number of electrons in the system.
Elementary charged excitations from this pseudospin polarized 
ground state correspond to the ground state configurations 
of the system with additional/missing flux $\pm 1$. 
At $\nu_n=1/q$, we study two types of excitations:
Laughlin's quasiholes (quasiparticles)\cite{QHE_text} and 
skyrmion quasiholes (quasiparticles). 
\cite{Sondhi_1993,Fertig_1994,Moon_1995,Rezayi_1991,Yoshioka_1998}
Laughlin's quasiholes (quasiparticles) 
correspond to pseudospin polarized excitations with $\pm 1$ flux, 
whose creation energy is given by  
\bea
\Delta_c^{\pm}=E(N_{\phi}\pm 1,P=1)-E(N_{\phi},P=1),
\eea
where $\pm$ represents quasiholes and quasiparticles, respectively,
and $P$ is the polarization ratio of the pseudospin, i.e. $P\equiv
 (N_K-N_{K'})/(N_{K}+N_{K'})$ with $N_K$ $(N_{K'})$ 
being the number of electrons in $K$ $(K')$ valley.

Skyrmion quasiholes (quasiparticles) 
correspond to pseudospin singlet excitations, and their 
creation energy is given by
\bea
\Delta_s^{\pm}=E(N_{\phi}\pm 1,P=0)-E(N_{\phi},P=1),
\eea
which could be smaller than $\Delta_c^{\pm}$.\cite{Sondhi_1993,Fertig_1994,Moon_1995,Rezayi_1991,Yoshioka_1998}

The activation energy, referred as the gap in the following, 
is given as a sum of these quasihole and quasiparticle energies,
$\Delta_c=\Delta_c^{+}+\Delta_c^{-}$ for pseudospin polarized 
(Laughlin-like)
excitations, and $\Delta_s=\Delta_s^{+}+\Delta_s^{-}$ for 
pseudospin unpolarized 
(skyrmion-like) 
excitations. 

At $\nu_n=2/5$ and 2/3, the pseudospin-polarized ground state and
unpolarized ground state compete each other. Unfavorably, in the
spherical geometry, these ground states realize at different
configurations (the number of electrons and the total flux). 
Precisely, the pseudospin-polarized $\nu_n=2/5$ state with $N_e$
electrons occurs when the total flux $N_{\phi}$ is given by
$N_{\phi}=(5/2)N_e-4$. On the other hand, the pseudospin-singlet
(unpolarized) state is realized at $N_{\phi}=(5/2)N_e-3$. Since the
finite size effects are different between these different
configurations, it is difficult to study the 
pseudospin-polarizability 
in the ground state in the sphere geometry. To avoid this difficulty, we
utilize the torus geometry in which two ground states with $P=0$ and
$P=1$ are realized in the same configuration: $N_{\phi}=(5/2)N_e$. We
use the sphere geometry to extrapolate the energy gaps to 
the thermodynamic limit.
The $\nu_n=2/3$ states are studied as well in the following section.
The elementally charge excitations at $\nu_n=2/3$ are obtained 
by adding or removing single electron in the system 
with $\pm 1$ flux.

\section{Results}
\subsection{$\nu_n=1$ and $1/3$ states} 
At $\nu_n=1$ integer fillings, the 
pseudospins of valley degrees of freedom 
in SLG and BLG are completely polarized by 
the exchange Coulomb interaction. This is 
essentially the same as the case
of usual quantum Hall systems with spin degrees
of freedom. 
The fully pseudospin polarized $P=1$ quasiparticle excitations 
from the above ground state need energy to the next higher LL, 
and such excitations have large gap compared with the
unpolarized $P=0$ or partially polarized $0<P<1$ excitations 
in the same LL. We therefore calculate the unpolarized and 
partially polarized excitations by the DMRG method, 
and study the pseudospin polarization 
of the lowest charge excitation and the gap 
in the thermodynamic limit.

\begin{figure}[t]
\begin{center}
\includegraphics[width=0.51\textwidth]{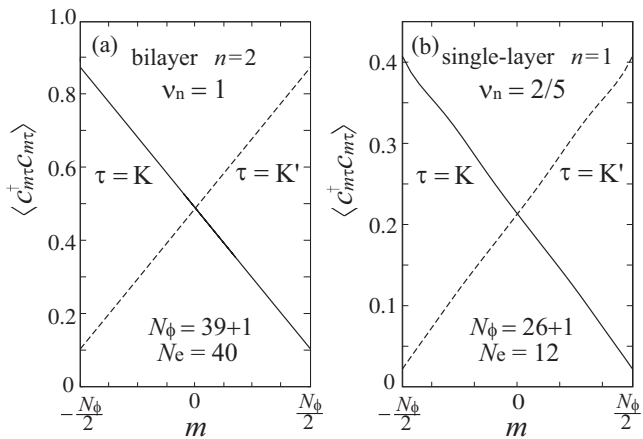}
\caption{
Expectation values, $\langle c_{m\tau}^{\dag}c_{m\tau}^{}\rangle$,
in the lowest pseudospin (valley) unpolarized excited state 
with extra one flux at (a) $\nu_2=1$ in BLG and 
at (b) $\nu_1=2/5$ in SLG. 
}
\label{figure2}
\end{center}
\end{figure}

The pseudospin polarization and the energy of the
lowest charge excitation are shown in Fig.~1 
for various sizes of system in spherical geometry.
This figure shows that the pseudospin unpolarized ($P=0$)
state is the lowest excited state when the system size is large enough
in SLG ($n=$ 0, 1, and 2) and BLG ($n=2$). 
The extrapolated value of the pseudospin unpolarized 
excitation gaps $\Delta_s$ in units of $e^2/(\epsilon \ell_B)$
are 0.63, 0.28 and 0.42 in the $n=0,1$ and 2 LLs of SLG,
and 0.48 in the $n=2$ LL of BLG.

In the $n=2$ LL, 
the lowest quasiparticle excitation in small systems 
($N_e  \stackrel{<}{_\sim} 20$)
has large pseudospin polarization $P=1-2/N_e$ both for SLG and BLG. 
These results show instability of the unpolarized $P=0$ excitations 
in small systems.
We find the first order transition in the lowest charge excited state 
when the number of electrons $N_e$ exceeds 18 in SLG and 24 in BLG.
This is consistent with the expectation that the unpolarized 
excitations are unstable in higher LLs
because of the long-range nature of the effective exchange 
interaction,
which increases energy of skyrmion-like  pseudospin unpolarized
state. 

To study the pseudospin structure in the unpolarized excited states, 
we compare our numerical results with the Hartree-Fock (HF) 
trial states of skyrmions.
The HF trial state of quasihole skyrmions at $\nu_n=1$ is 
written in the form:\cite{QHF_review} 
\bea
 |\Psi_{sk}\rangle
=\prod_{m=-N_{\phi}/2}^{N_{\phi}/2}
[{\alpha}_m^{}
c_{m K}^{\dag}+{\beta}_m^{}c_{m+1 K'}^{\dag}]|0\rangle, 
\eea
where
$\langle c_{mK}^{\dag}c_{mK}^{\ }\rangle=|\alpha_m|^2$ and
$\langle c_{mK'}^{\dag}c_{mK'}^{\ }\rangle=|\beta_{m-1}|^2$.
we have calculated the expectation values 
$\langle c_{m\tau}^{\dag}c_{m\tau}^{\ }\rangle$
from the wave function obtained in the present DMRG study.
The results for $\nu_2=1$ in BLG indicate they are approximately given by 
$\langle c_{m\tau}^{\dag}c_{m\tau}^{\ }\rangle=1/2 \mp m/N_{\phi}$
for $\tau=K$ and $\tau=K'$, respectively as shown in Fig.~2 (a).
Similar results are also obtained for $\nu_n=1$ and $1/3$ 
in the $n=0$ and 1 LLs of SLG, that shows the pseudospin 
unpolarized
elementally charge excitations in BLG at $\nu_2=1$
are pseudospin (valley) skyrmions.

\begin{figure}[t]
\begin{center}
\includegraphics[width=0.47\textwidth]{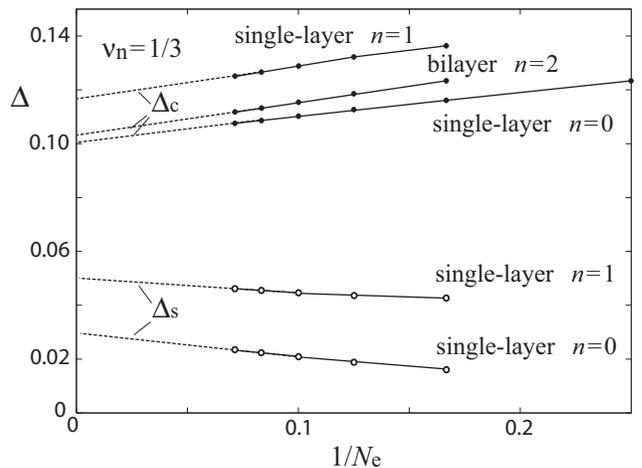}
\caption{
The pseudospin (valley) polarized excitation gap $\Delta_c$
and the pseudospin (valley) unpolarized (skyrmion)
excitation gap $\Delta_s$
at $\nu_n=1/3$ in SLG ($n=0$ and 1 LLs) and in BLG ($n=2$ LL).
}
\label{figure3}
\end{center}
\end{figure}

The elementally charge excitation gaps at fractional fillings
$\nu_n=1/3$ are shown in Fig.3.
The pseudospin (valley) polarized $P=1$ excitation 
gap $\Delta_c$ for $n=0$ SLG is 0.101 in units of 
$e^2/(\epsilon \ell_B)$ in the thermodynamic limit, 
which is in a good agreement with the previous work.\cite{Morf_2002}
In the $n=1$ LL of SLG, $\Delta_c$ is 0.115,
which is larger than 0.101 in the $n=0$ LL of SLG. 
In BLG, $\Delta_c=0.103$ in the $n=2$ LL,
which is also slightly larger than that in the $n=0$ LL of SLG.
This enhancement of the $\Delta_c$ in higher LLs is consistent
with the increase of the difference of Haldane's pseudopotentials 
between $m=1$ and 3; $V_1^n-V_3^n$ is 0.168 
in units of $e^2/(\epsilon \ell_B)$ 
for BLG in the $n=2$ LL and 0.198 for SLG 
in the $n=1$ LL, while it is 0.166 
for SLG in the $n=0$ LL.
  
As seen in conventional 2DEGs, $\nu=1/3$ (valley or spin) unpolarized excited
states can be lower in energy than the polarized excited states.  
Indeed, as shown in Fig.~3, the unpolarized excitation gap 
$\Delta_s$ in SLG is 0.05 in the $n=0$ LL and 0.03 in the $n=1$ LL, 
which are much smaller than the polarized excitation gap $\Delta_c$
in the $n=0$ and 1 LLs, respectively. 
In the case of BLG, however, we find that valley unpolarized excited 
states have higher energy than polarized excited states. 
Consequently, the activation 
energy at $\nu_2=1/3$ in BLG is 2-3 times larger than that in SLG,
that indicates an advantage for the observation of fractional
QHE in BLG.

\begin{figure}[!t]
\begin{center}
\includegraphics[width=0.48\textwidth]{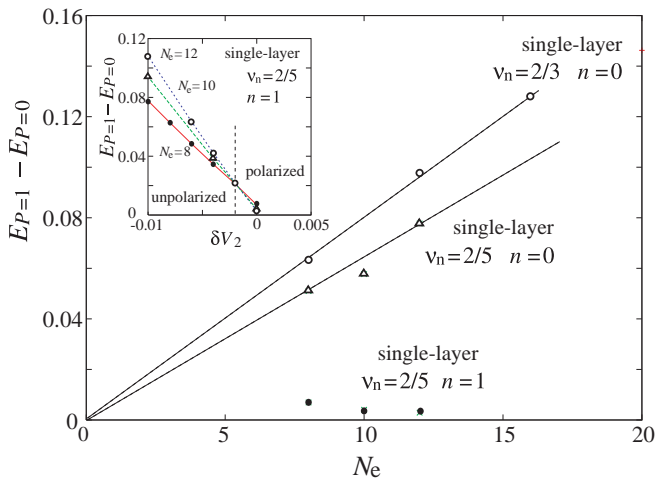}
\caption{Size dependence of the energy difference
between the fully pseudospin polarized $P=1$ state $E_{P=1}$ and 
the unpolarized $P=0$ states $E_{P=0}$ at 
$\nu_n=2/5$ in SLG in a torus geometry.
Inset shows size dependence and $V_2$ dependence of $E_{P=1}-E_{P=0}$
at $\nu_n=2/5$ in SLG with $n=1$.}
\label{figure4}
\end{center}
\end{figure}

\subsection{$\nu_n=2/5$ and $\nu_n=2/3$ states}

According to the composite fermion theory\cite{Jain_1989},  
$\nu=m/(1+2m)$ fractional quantum Hall states are 
mapped onto the $\nu_{\rm eff}=m$ integer quantum Hall states.
In contrast to the case of $\nu=1/3$, 
where the ground state is mapped onto the $n=1$ pseudospin 
{\it polarized} integer quantum Hall state,
$\nu=2/5$ fractional quantum Hall state is mapped onto the $n=2$
integer quantum Hall state, 
where the lowest LL is doubly occupied by pseudospin-up 
and pseudospin-down electrons. 
We therefore expect pseudospin unpolarized ground state at $\nu_n=2/5$. 
Here we calculate the ground state
pseudospin polarization 
of SLG and BLG at $\nu_n=2/5$ to see whether 
this naive expectation is correct even in graphene.

The pseudospin-polarization in the ground state is studied by
calculating the polarization energy 
$E_{P=1}-E_{P=0}$
which is the energy difference between the polarized $P=1$ state 
and the unpolarized $P=0$ state.
This energy difference is plotted in Fig.~4 
for the $n=0$ and $1$ LLs in SLG as a function of $N_e$. 
This figure clearly shows that the ground state pseudospin
at $\nu_n=2/5$ in the $n=0$ LL of SLG is unpolarized, 
because $E_{P=1}-E_{P=0}$ increases
with the increase in the number of electrons $N_e$.
This is consistent with the prediction of the
composite fermion theory.
On the other hand, in the 
$n=1$ LL of SLG, $E_{P=1}-E_{P=0}$ decreases with the increase in $N_e$.
This behavior suggests the polarized ground state.

\begin{figure}[!t]
\begin{center}
\includegraphics[width=0.48\textwidth]{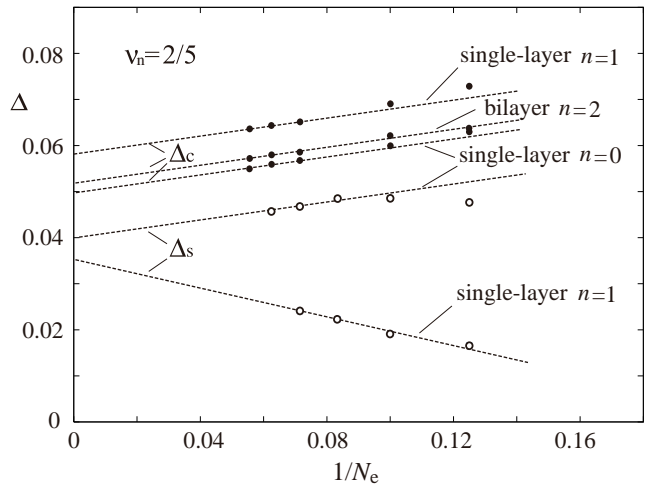}
\caption{The pseudospin (valley) polarized excitation gap $\Delta_c$
and the pseudospin (valley) unpolarized 
excitation gap $\Delta_s$
at $\nu_n=2/5$ in $n=0$ and 1 LLs of SLG and in the $n=2$ LL of
BLG.
}
\label{figure5}
\end{center}
\end{figure}

To confirm the polarized ground state at $\nu_n=2/5$
in the $n=1$ LL of SLG, we slightly change the Haldane's 
pseudo potential $V_2$, which acts only for electron 
pairs whose relative angular momentum $m$ is 2.
Since anti-symmetrized wave function of polarized 
pseudospin state does not contain electron 
pairs with $m=2$, only the energy of pseudospin 
polarized state is independent of $V_2$.
We therefore systematically control the energy difference 
between $E_{P=1}$ and $E_{P=0}$ by changing $V_2$.
The $V_2$-dependence of $E_{P=1}-E_{P=0}$ 
is shown in the inset of Fig.4 
as a function of $\delta V_2=V_2-V_2^{\rm SLG(n=1)}$
where $V_2^{\rm SLG(n=1)}$ corresponds to the original
$V_2$ in the $n=1$ LL of SLG. 
This $V_2$-dependence shows systematic change in 
the size dependence of $E_{P=1}-E_{P=0}$ at $\delta V_2=-0.002$.
Since $\delta V_2$ is negative at the transition, where 
$E_{P=1}-E_{P=0}=0$ in the thermodynamic limit, 
the pseudospins in the ground state at $\nu_n=2/5$ 
in $n=1$ LL of SLG is fully polarized in large systems.
Similar analysis on the size dependence of the polarization 
energy also shows that the pseudospins in the ground state 
of $n=2$ LL of BLG are fully polarized at $\nu_n=2/5$.

The elementally charge excitation energies at $\nu_n=2/5$
are presented in Fig.~5.
In the case of $n=0$ LL of SLG, 
 the pseudospin (valley) polarized excitation gap
$\Delta_c$ is calculated 
supposing pseudospins in the ground state are fully polarized,
although they are unpolarized in the true ground state.
The extrapolated value of $\Delta_c$ in the $n=0$ LL of SLG 
is then $0.05$ $e^2/(\epsilon \ell_B)$ 
in a good agreement with the previous work.\cite{Morf_2002}
In the $n=1$ LL of SLG, the pseudospins in the ground state 
are fully polarized and $\Delta_c$ is 0.059. In BLG, $\Delta_c=0.052$,
which is also slightly larger than $\Delta_c$ in the $n=0$ LL of SLG.
The largest $\Delta_c$ at $\nu_n=2/5$ is obtained in the $n=1$ LL of SLG.
This feature is the same as the case of $\nu_n=1/3$ shown in Fig.~3.

The pseudospin (valley) unpolarized excitation gap $\Delta_s$  at 
$\nu_n=2/5$ are also shown in Fig.~5.
The $\Delta_s$ from the unpolarized ground state in the $n=0$ LL 
of SLG is 0.04 $e^2/(\epsilon \ell_B)$, 
and the $\Delta_s$ from the polarized ground state in the $n=1$ LL 
of SLG 0.035 $e^2/(\epsilon \ell_B)$.
These $\Delta_s$ are smaller than $\Delta_c$
similarly to the case of $\nu_n=1/3$.
In the $n=2$ LL of BLG, $\Delta_s$ is larger than $\Delta_c$.
The lowest gap in BLG is then given by $\Delta_c$, which is 0.052 
$e^2/(\epsilon \ell_B)$.

In the $n=1$ LL of SLG, the ground state is pseudospin
polarized while the lowest charge excited state is
pseudospin unpolarized. The pseudospin structure in 
the unpolarized excited states at $\nu_1=2/5$ is 
shown in Fig.~2 (b),
which indicates elementally charge excitations are 
characterized by valley skyrmions.
The two-particle correlation functions $g_{\tau\tau'}(r)$
\cite{Haldane_1983,QHE_text,Yoshioka_1998}
at $\nu_1=2/5$ are presented in Fig.~6, which 
show that the correlation function between the electrons 
in the same valley $g_{KK}(r)$ in the unpolarized quasihole state 
has peak structure around the origin while 
the correlation function between the electrons 
in different valleys $g_{KK'}(r)$ has the maximum at the opposite 
side on the sphere.\cite{Shibata_2008,Yoshioka_1998}
These results indicate the elementally charge excitations at
$\nu_n=2/5$ in the $n=1$ LL of SLG are valley-pseudospin textures 
similar to skyrmion excitations in the quantum Hall ferromagnetic states
at $\nu_n=1$ and $\nu_n=1/3$, although their origin and detailed
properties are not obvious. 

\begin{figure}[t]
\begin{center}
\includegraphics[width=0.48\textwidth]{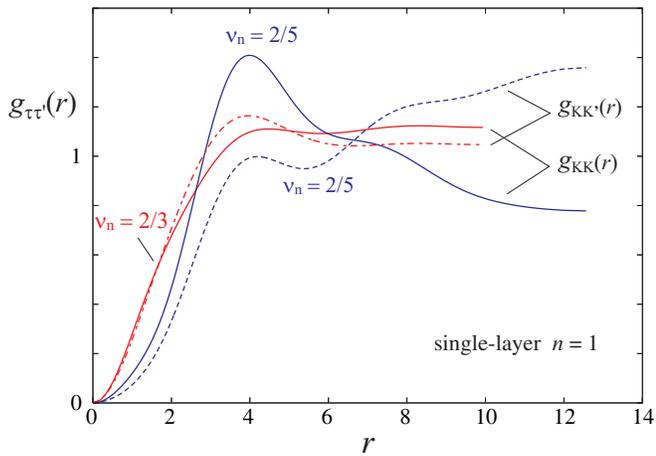}
\caption{
(Color online) Two-particle correlation functions $g_{\tau\tau'}$ of quasihole 
skyrmion state at $\nu_n=2/5$ and  
quasiparticle state at $\nu_n=2/3$ in the $n=1$ LL of SLG. 
$r$ is in units of $\ell_B$.}
\label{figure7}
\end{center}
\end{figure}

We finally study the valley-polarization in the ground state 
and the elementally excitations at $\nu_n=2/3$. 
Similarly to the case of the $\nu_n=2/5$ fractional quantum Hall state,
the $\nu_n=2/3$ fractional quantum Hall state is mapped onto $n=2$
integer quantum Hall state within a mean-field analysis
of the composite fermion theory.
Thus the pseudospin unpolarized ground state is expected.
Indeed, the ground state at $\nu_n=2/3$ in the $n=0$ LL of SLG 
is pseudospin unpolarized.
However, in the $n=1$ LL of SLG and the $n=2$ LL of BLG, 
the size dependence of the polarization energy shows 
the pseudospins are fully polarized in the ground state.

The charge excitation energies from the ground states at 
$\nu_n=2/3$ are presented in Fig.~7.
The polarized excitation gap $\Delta_c$ in the 
$n=0$ LL of SLG is calculated from the energy difference 
between the fully polarized ground state and the
fully polarized charge excited state,
although the true ground state is the unpolarized state.
The extrapolated value of $\Delta_c$ in the thermodynamic limit 
is then $0.101$ $e^2/(\epsilon \ell_B)$ in the $n=0$ LL of SLG, 
which is the same as $\Delta_c$ at $\nu=1/3$
because of the particle-hole symmetry of the 
single component quantum Hall system.
In the $n=1$ LL of SLG, the pseudospins in the ground state 
are polarized and 
$\Delta_c$ is 0.115 in the thermodynamic limit.
In BLG, $\Delta_c=0.103$,
which is also the same as $\Delta_c$ at $\nu_n=1/3$.

The valley unpolarized excitation gap $\Delta_s$ from the 
unpolarized true ground state in the $n=0$ LL of SLG
is 0.08 $e^2/(\epsilon \ell_B)$.
Similarly to the case of $\nu=1/3$, 
$\Delta_s$ is smaller than $\Delta_c$.
In the $n=1$ LL of SLG, we find $\Delta_s$ is 0.10 in the 
thermodynamic limit, which is also smaller than $\Delta_c$.
The valley unpolarized elementally charge excitations at
$\nu_n=2/3$ in the $n=1$ LL of SLG are not skyrmion-like
pseudospin-textured state, in contrast to the case of $\nu=2/5$ 
in the $n=1$ LL of SLG as shown in Fig.~6, where 
short-range correlation $g_{KK}$ is smaller than 
$g_{KK'}$ for $\nu_n=2/3$.

\begin{figure}[t]
\begin{center}
\includegraphics[width=0.45\textwidth]{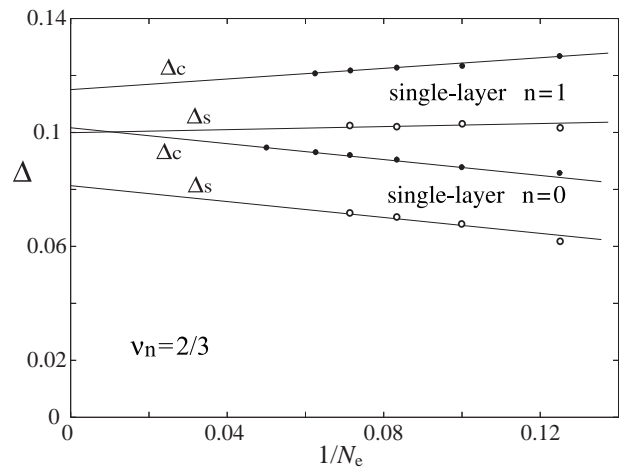}
\caption{The pseudospin (valley) polarized excitation gap $\Delta_c$
and the pseudospin (valley) unpolarized 
excitation gap $\Delta_s$ at $\nu_n=2/3$ in $n=0$ and 1 LLs
of SLG.
}
\label{figure6}
\end{center}
\end{figure}

\begin{table}[t]
\caption{The pseudospin (valley) polarization in the 
ground state and the lowest charge excited state, and 
the unpolarized excitation gap $\Delta_s$ and the polarized
excitation gap $\Delta_c$ extrapolated to the thermodynamic limit
at the effective filling $\nu_n$ in the $n$th LL
of single-layer graphene (SLG) and bilayer graphene (BLG).
}
\begin{tabular}{ccccccc}
\hline
$\nu_n$& & LL & ground state  & excited state& $\Delta_s$ & $\Delta_c$\\
\hline
1  &SLG &0 & polarized & skyrmion & 0.63 &\\
1  &SLG &1 & polarized & skyrmion & 0.28 &\\
1  &SLG &2 & polarized & skyrmion & 0.42 &\\
1  &BLG &2 & polarized & skyrmion & 0.48 &\\
\hline
1/3&SLG &0 & polarized & skyrmion & 0.05 & 0.101\\
1/3&SLG &1 & polarized & skyrmion & 0.03 & 0.115\\
1/3&BLG &2 & polarized & polarized & &  0.103 \\
\hline
2/5&SLG &0 & unpolarized &  unpolarized & 0.04& (0.050)\cite{chargegap}\\
2/5&SLG &1 & polarized & skyrmion & 0.035& 0.058 \\
2/5&BLG &2 & polarized & polarized & & 0.052\\
\hline
2/3&SLG &0 & unpolarized &  unpolarized & 0.08 & (0.101)\cite{chargegap}\\
2/3&SLG &1 & polarized & unpolarized & 0.10 & 0.115\\
2/3&BLG &2 & polarized & polarized & & 0.103\\
\hline
\end{tabular}
\end{table}

\section{ Discussion}

Our DMRG calculation confirms 
various types of quantum Hall states in graphene 
at $\nu_n=1, 1/3, 2/5$ and $2/3$ in the $n=0$ and 1 LLs of SLG 
and in the $n=2$ LL of BLG.
These results are summarized in Table I, where 
the pseudospin polarizations 
for the ground state and the lowest charge excited state
are listed with the elementally charge
excitation energies $\Delta_c$ and $\Delta_s$.
The elementally charge excitations are obtained by
increasing or decreasing the flux quantum number $N_{\phi}$ by 1
for $\nu=1$ and $1/3, 2/5$, and 
by increasing or decreasing the flux quantum number with 
adding or removing single electron in the system for $\nu=2/3$.
We have studied both (a) pseudospin polarized excitations (Laughlin's
quasiholes and quasiparticles) and (b) pseudospin unpolarized 
excitations (quasihole skyrmions and quasiparticle skyrmions
at $\nu_n=1$ and $1/3$ in the $n=0$ and 1 LLs of SLG, 
and at $\nu_n=1$ in the $n=2$ LL of BLG). 

The activation energies obtained in finite systems are extrapolated to
the thermodynamic limit, which give theoretical predictions for future
experimental studies of the fractional quantum Hall states in graphene.  
Our results show that the gaps ($\Delta_s\simeq 0.1 e^2/\ell_B$) at
$1/3$ and $2/3$ effective fillings  
 in the excited LL in bilayer graphene are larger than those in
 conventional quantum well and single layer graphene. Therefore bilayer
 graphene is a good candidate for future experimental observation of the
 fractional quantum Hall effects. 

\section*{Acknowledgment}

The present work is supported by Grand-in-Aid No. 18684012 from MEXT Japan.


\begin{thebibliography}{99}
\bibitem{chargegap} $\Delta_c$ for $\nu_n=2/5$ and 2/3 in the $n=0$ LL of SLG 
is the gap from the lowest polarized state.
\end{thebibliography}
\end{document}